\documentclass[conference]{IEEEtran}
\IEEEoverridecommandlockouts
\usepackage{cite}
\usepackage{amsmath,amssymb,amsfonts}
\usepackage{mathtools}
\usepackage{algorithmic}
\usepackage{textcomp}
\usepackage{xcolor}
\def\BibTeX{{\rm B\kern-.05em{\sc i\kern-.025em b}\kern-.08em
    T\kern-.1667em\lower.7ex\hbox{E}\kern-.125emX}}

\ifCLASSINFOpdf
\usepackage[pdftex]{graphicx}
\graphicspath{{./images/}}
\else
\fi

\usepackage{subfig}

\DeclarePairedDelimiter\ceil{\lceil}{\rceil}
\DeclarePairedDelimiter\floor{\lfloor}{\rfloor}

\newcommand{\eabsn}[2]{\left|  #1 \right| ^{#2}}

\newcommand{\eceil}[1]{\ceil*{#1}}
\newcommand{\efloor}[1]{\floor*{#1}}

\newcommand{\emaxs}[2]{\max_{#1}\left(#2\right) }

\newcommand{\eds}[1]{\left[  #1 \right] }
\newcommand{\ecs}[1]{\left(   #1 \right)  }

\def\mydate{\leavevmode\hbox{\the\year/\twodigits\month/\twodigits\day}}
\def\twodigits#1{\ifnum#1<10 0\fi\the#1}

\newcommand{\eAddFig}[4]{
	\begin{figure}[!t]
		\centering
		\includegraphics[width=#2\linewidth]{#1}
		\vspace{-3mm}
		\caption{#4}
		\label{#3}
	\end{figure}
}

\newcommand{\eAddTwoColFig}[4]{
	\begin{figure*}[!t]
		\centering
		\includegraphics[width=#2\linewidth]{#1}
		\vspace{-2mm}
		\caption{#4}
		\label{#3}
	\end{figure*}
}

\usepackage[acronym]{glossaries}

\newacronym{soa}{SoA}{state of the art}
\newacronym{cep}{CEP}{channel estimation procedure}
\newacronym{ts}{TS}{training sequence}

\newacronym{gps}{GPS}{Global Positioning System}

\newacronym{3gpp}{3GPP}{3rd Generation Partnership Project}
\newacronym{5g}{5G}{Fifth Generation}
\newacronym{6g}{6G}{Sixth Generation}
\newacronym{embb}{eMBB}{enhanced mobile broadband}
\newacronym{urllc}{URLLC}{ultra reliable and low-latency communications}
\newacronym{ritm}{RITM}{RIS-In-The-Middle}
\newacronym{dris}{D-RIS}{defensive-RIS}

\newacronym{toa}{ToA}{time of arrival}
\newacronym{aoa}{AoA}{angle of arrival}
\newacronym{aod}{AoD}{angle of departure}
\newacronym{rss}{RSS}{received signal strength}
\newacronym{crlb}{CRLB}{Cramér-Rao lower bound}

\newacronym{ssb}{SSB}{synchronization signal block}
\newacronym{pss}{PSS}{primary synchronization signal}
\newacronym{sss}{SSB}{secondary synchronization signal}
\newacronym{pbch}{PBCH}{physical broadcast channel}
\newacronym{prach}{PRACH}{physical random access channel}
\newacronym{dmrs}{DM-RS}{demodulation reference signals}

\newacronym{mmW}{mmWave}{millimetre waves}
\newacronym{cmW}{cmWave}{centimetre waves}
\newacronym{dmW}{dmWave}{decimetre waves}
\newacronym{bmp}{BMP}{beam-management procedure}
\newacronym{iap}{IAP}{initial access procedure}

\newacronym{ris}{RIS}{reconfigurable intelligent surface}
\newacronym{fsm}{FSM}{frequency shifter meta-surface}
\newacronym{comp}{CoMP}{coordinated multi-point}

\newacronym{bs}{BS}{base station}
\newacronym{ue}{UE}{user equipment}
\newacronym{tdl}{TDL}{tapped delay line}
\newacronym{umi}{UMi}{urban micro-cell}

\newacronym{tdm}{TDM}{time-division multiplexing}
\newacronym{fdm}{FDM}{frequency-division multiplexing}
\newacronym{tdd}{TDD}{time-division duplexing}
\newacronym{fdd}{FDD}{frequency-division duplexing}
\newacronym{ul}{UL}{uplink}
\newacronym{dl}{DL}{downlink}
\newacronym{csi}{CSI}{channel state information}

\newacronym{upa}{UPA}{uniform planar array}
\newacronym{ula}{ULA}{uniform linear array}

\newacronym{dft}{DFT}{discrete Fourier transform}
\newacronym{idft}{IDFT}{inverse discrete Fourier transform}

\newacronym{cp}{CP}{cyclic prefix}
\newacronym{ofdm}{OFDM}{orthogonal frequency division multiplexing}

\newacronym{tr}{TR}{tone reservation}

\newacronym{pa}{PA}{power amplifier}
\newacronym{sspa}{SSPA}{solid state power amplifier}

\newacronym{ls}{LS}{Least-Squares}
\newacronym{clt}{CLT}{Central Limit Theorem}

\newacronym{cordic}{CORDIC}{coordinate rotation digital computer}
\newacronym{qcqp}{QCQP}{quadratically constrained quadratic program}

\newacronym{psam}{PSAM}{pilot symbol assisted modulation}
\newacronym{st}{ST}{superimposed training}
\newacronym{htst}{HT-ST}{hollow tone-aided superimposed training}
\newacronym{tlst}{TLST}{two layer superimposed training}

\newacronym{mrt}{MRT}{maximum ratio transmission}
\newacronym{mrc}{MRC}{maximum ratio combining}
\newacronym{zf}{ZF}{zero-forcing}
\newacronym{mimo}{MIMO}{multiple-input multiple-output}

\newacronym{qam}{QAM}{quadrature amplitude modulation}

\newacronym{los}{LOS}{line-of-sight}
\newacronym{nlos}{NLOS}{non-line-of-sight}
\newacronym{awgn}{AWGN}{additive white Gaussian noise}
\newacronym{papr}{PAPR}{peak-to-average power ratio}
\newacronym{snr}{SNR}{signal to noise ratio}
\newacronym{sinr}{SINR}{signal to interference and noise ratio}
\newacronym{ber}{BER}{bit error rate}
\newacronym{ser}{SER}{symbol error rate}
\newacronym{mse}{MSE}{mean squared error}
\newacronym{isi}{ISI}{inter-symbol interference}
\newacronym{ici}{ICI}{inter-carrier interference}
\newacronym{asr}{ASR}{achievable secrecy rate}

\newacronym{ols}{OLS}{optical label switching}
\newacronym{lan}{LANs}{local area networks}
\newacronym{ip}{IP}{Internet Protocol}
\newacronym{qp}{QP}{Quantum Protocol}
\newacronym{stp}{STP}{Spanning Tree Protocol}
\newacronym{qstp}{Q-STP}{quantum-STP}
\newacronym{mac}{MAC}{medium access control}
\newacronym{icmp}{ICMP}{Internet Control Message Protocol}
\newacronym{arp}{ARP}{Address Resolution Protocol}
\newacronym{crc}{CRC}{cyclic redundancy check}

\newacronym{qkd}{QKD}{quantum key distribution }

\begin{document}

\title{Sequential Entanglement-Swapping assisted by Quantum Protocol over Ethernet Networks\\

\thanks{This work has been funded by the Villum Investigator Grant "WATER" from the Velux Foundation (Denmark), and Centers of Excellence Grant "CLASSIQUE" from Denmark's Basic Research Foundation.}

\author{\IEEEauthorblockN{Kun Chen-Hu, Kristian S. Jensen, and Petar Popovski}
\IEEEauthorblockA{\textit{Department of Electronic Systems, Aalborg University, Denmark} \\
E-mails: \{kchenhu, ksjen, petarp\}@es.aau.dk}}

}

\maketitle

\begin{abstract}
The integration of quantum communication protocols over Ethernet networks is proposed, showing the potential of combining classical and quantum technologies for efficient, scalable quantum networking. By leveraging the inherent strengths of Ethernet, such as addressing, MAC layer functionality, and scalability; we propose a practical framework to support the rigorous requirements of quantum communication. Some novel protocols given in this study enable reliable end-to-end quantum entanglement over Ethernet, ensuring the adaptability needed for implementing a stable quantum internet. Detailed time-delay analyses confirm that our protocols offer superior performance compared to existing methods, with total time delay kept within the decoherence threshold of qubits. These results suggest that our approach is well-suited for deployment in realistic environments, meeting both the immediate needs of quantum networking and laying the groundwork for future advances in data exchange and quantum computational capabilities.
\end{abstract}

\begin{IEEEkeywords}
delay, Ethernet, protocol, quantum, swapping,
\end{IEEEkeywords}

\section{Introduction}

The quantum internet is the next-generation communication system, leveraging the principles of quantum mechanics to create ultra-secure, high-speed networks through quantum entanglement and \acrfull{qkd}. Unlike traditional internet systems, qubits enable more robust and secure information transfer, making it an ideal platform for secure communications. Moreover, the quantum internet also supports distributed quantum computing, enabling collaborative quantum processing across multiple nodes, and networked quantum sensing, which enhances precision in applications like metrology and environmental monitoring. These advancements could revolutionize data sharing for sectors where privacy and accuracy are paramount, offering a foundation for global quantum communication and laying the groundwork for future quantum cloud computing systems.

\eAddTwoColFig{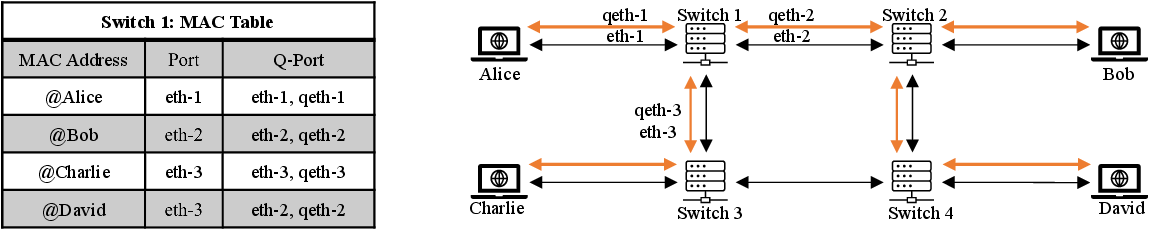}{0.93}{fig1switches}{Illustrative example of the proposed wired or wireless quantum network coexisting with the traditional Ethernet network, built by $U_{t}=4$ devices and $S_{t}=4$ switches. The MAC table of the switch $\#1$ shows that both the classical Ethernet port (eth) and the new quantum port (qeth) are accounted for addressing.}

Quantum networks depend on classical ones for two key functions: transferring supporting data, such as the classical bits needed for quantum teleportation, and transmitting metadata required to send and interpret quantum data accurately. This reliance stems from the unique properties of qubits, which cannot be copied or directly observed without disrupting their quantum state, as dictated by the no-cloning theorem \cite{clon}. Classical networks handle the coordination and switching of qubits by tracking entangled pairs and establishing secure channels between nodes, ensuring that quantum data is delivered to the correct destination without interference. This integration of classical and quantum networks is fundamental to realizing a fully operational quantum internet.


Several studies in the literature have contributed to the development of a quantum internet protocol stack \cite{qp1, qp2, qp3}. These works focus on defining a layered protocol stack that enables efficient deployment of essential quantum operations, such as entanglement swapping and purification. However, the convergence between classical and quantum networks remains insufficiently addressed. As noted in \cite{qsurv1, qsurv2}, effective signalling mechanisms are crucial for managing multiple repeaters in the network, including discovery, forwarding, and switching operations. Additionally, a robust \acrfull{mac} layer is required to manage resource competition and address the various entities involved in transmission \cite{qmag1}. Other works have focused on entanglement routing \cite{qrou1, qrou2, qrou3}, developing optimal routing algorithms for specific network topologies while assuming the presence of a centralized unit. However, these approaches overlook the need for signalling mechanisms in realistic network conditions, and they significantly scale down the network size, limiting practical applicability \cite{qrou3}. Recently, \acrfull{ols} \cite{ols} has been proposed as a quantum wrapper to manage quantum transmissions over optical fibres \cite{qwrap}. While OLS leverages existing protocols, it falls short of addressing MAC-related challenges, as it operates exclusively over point-to-point links, limiting scalability in multi-user scenarios.

Ethernet \cite{std-eth} is one of the most widely used and reliable technologies for creating distributed \acrfull{lan}. In contrast to the global internet, which operates as a \textit{network of networks} connecting diverse domains and infrastructures, Ethernet is specifically designed for efficient networking within locally controlled topologies, while simultaneously contributing to the broader internet. Its localized design allows Ethernet to function as a self-contained network within the global framework, enabling local interactions with reduced overhead, lower latency, and enhanced performance \cite{eth-lat}. Building on these advantages, this work proposes, for the first time to our knowledge, the deployment of the quantum internet over Ethernet to support locally distributed quantum information processing, such as computing, secure key generation, and sensing. This approach leverages the strengths of Ethernet, including scalable networks, to support crucial functions such as switching, addressing, and link management, which are vital for reliable qubit transmission. A novel \acrfull{qp} is introduced to facilitate entanglement swapping among network entities in an existing Ethernet network. Furthermore, the signalling requirements for synchronization and a comprehensive time-delay assessment are analysed to evaluate the feasibility of the proposed protocol.

An example of the proposed wired or wireless distributed quantum network is depicted in Fig. \ref{fig1switches}. Unlike \cite{qwrap}, the switches manage their switching table with two types of interfaces for addressing purposes: the traditional Ethernet ports and the new quantum ports (qeth). Despite the quantum channel needs the coordination of the classical one, both links can be independently managed, they do not require accurate synchronization to each other.


The main contributions are summarized as follows:
\begin{itemize}
	\item An Ethernet network must be self-managed and ready to support the integrated transmission of both bits and qubits. Following \acrfull{stp}\cite{stp,rfc-stp}, a \acrfull{qstp} is proposed to eliminate network loops by jointly considering the classical and quantum links. Then, a Discovery Protocol identifies a viable path for both classical and quantum data from the source to the destination, updating the switching tables. Later, a Path Establishment Protocol coordinates the synchronization of a series of switches to perform entanglement swapping, establishing a virtual circuit to ensure consistent data flow between two users. Any modifications made by the network are promptly communicated, ensuring that data adheres strictly to the predefined path, and maintaining the integrity of the connection.
	
	\item The QP is presented to establish end-to-end entanglement between two users. The entanglement swapping is performed sequentially, progressing from the edge switches directly serving the users toward the central one in the route. Given the possibility of failure in the entanglement-swapping process, the proposed protocol includes mechanisms to notify all involved nodes and restart the process as needed, ensuring that all switches in the path are successfully entangled. A key aspect of this approach is that the quantum channel is only activated for qubit transmission once both endpoints successfully negotiate via the classical (handshake) channel. As a result, the quantum channel inherently benefits from well-established MAC and enhances its stability and efficiency in a multi-user environment.
	
	\item The time delay of each proposed protocol is analytically derived and compared with the approach in \cite{qwrap}, which does not allow for multi-user access. Our analysis indicates that the proposed protocol achieves lower delay, as qubits are transmitted only after a successful handshake via the classical channel, unlike \cite{qwrap}. This approach not only enhances robustness against qubit decoherence but also supports scalability as quantum technologies advance. Therefore, implementing QP over Ethernet offers a practical and forward-compatible solution for developing a scalable and resilient quantum internet.

\end{itemize}

The remainder of the paper is organized as follows. 
Section \ref{sec:qp} presents the idea of deploying the QP over Ethernet, specifying the content of a frame. 
Section \ref{sec:estab} describes some protocols required to set up the network before starting the entanglement procedure. 
Section \ref{sec:qpeth} addresses the end-to-end entanglement procedure and its time delay analysis.
Section \ref{sec:num} shows several numerical results for the proposed scheme, providing an assessment of the attained system performance. 
Finally, in Section \ref{sec:conclusion}, the conclusions are disclosed.

\section{Quantum Protocol over Ethernet}
\label{sec:qp}


Ethernet is an excellent option for implementing classical communications to manage distributed quantum computing within controlled topologies. Its high-speed transmission and low-latency performance \cite{eth-lat} are critical for coordinating real-time quantum operations. Moreover, its scalability supports growing data demands and the seamless integration of new quantum nodes as networks expand. These features make Ethernet a robust and adaptable foundation for the quantum internet, capable of meeting the complex requirements of secure and scalable quantum communication systems.

\subsection{Proposed System Architecture}

Let us consider a classical wired or wireless local area network (LAN) with topology described by a graph $G\ecs{\mathcal{U}_{t} \cup \mathcal{S}_{t}, \mathcal{L}_{c}}$ where $\mathcal{U}_{t}$ and $\mathcal{S}_{t}$ are the set that contains the total number of user devices and switches, respectively, and $\mathcal{L}_{c}$ corresponds to the set of classical communication links among the $\eabsn{\mathcal{U}_{t}}{} = U_{t}$ users and $\eabsn{\mathcal{S}_{t}}{} = S_{t}$ switches, which must satisfy that none of the users can be directly connected ($\ecs{u,u'} \notin \mathcal{L}_{c},  \forall u,u' \in \mathcal{U}_{t},$). On top of this classical network, let us define a quantum network with topology described by a graph $G\ecs{\mathcal{U}_{t} \cup \mathcal{S}_{t}, \mathcal{L}_{q}}$ where $\mathcal{L}_{q}$ is denoting the set of quantum links and it must also satisfy that ($\ecs{u,u'} \notin \mathcal{L}_{q},  \forall u,u' \in \mathcal{U}_{t},$).

The main objective of the proposed system architecture is to take advantage of an existing Ethernet network, which provides a classical communication service ($\mathcal{L}_{c}$), to deploy new quantum services. The existing Ethernet network is also responsible for managing the transmission of the qubits in the quantum links enabling an efficient end-to-end entanglement. The topology of the two networks does not have to be the same ($\mathcal{L}_{c} \neq \mathcal{L}_{q}$) since some switches may not be equipped with enough quantum links. In Fig. \ref{fig1switches}, the switches $\#3$ and $\#4$ are not connected by a quantum link. If Charlie and David would like to share qubits, it should be performed via switches $\#1$ and $\#2$. However, the qubits cannot be sent as headers like classical bits, since they cannot be observed, copied and retransmitted due to the no-cloning theorem \cite{clon}. A tuple of one classical and quantum link must be considered for deploying any quantum service since the latter relies on the former for transmitting its associated headers. From a quantum perspective, the topology of the two networks must satisfy
\begin{equation}\label{eq:linkrest}
	\ecs{u,s} \in \mathcal{L}_{c} \cap \mathcal{L}_{q}, \quad u \in \mathcal{U}_{t}, s \in \mathcal{S}_{t}  \quad \text{or} \quad u, s \in \mathcal{S}_{t},
\end{equation}
where it points out that the topology of the quantum links and those classical managing links must be the same.



\subsection{Proposed Transmission Mode}

\begin{figure}[!t]
	\centering
	\subfloat[Proposed scheme.]
	{\includegraphics[width=0.48\linewidth]{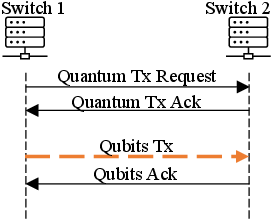}%
				\label{fig:fig21transprop}}
	\hfil
	\subfloat[Reference case \cite{qwrap}.]
	{\includegraphics[width=0.48\linewidth]{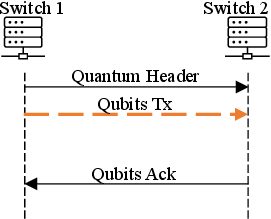}%
				\label{fig:fig21transref}}
	\caption{Proposed transmission mode for managing a quantum packet transmission based on performing a handshake in the classical channel, and its comparison to the reference case.}
	\label{fig:range}
\end{figure}

To reduce the time delay in transmitting a quantum packet, the transmission mode of both links must be jointly considered. For a given tuple of classical and quantum links, it is proposed to perform a handshake in the classical channel before transmitting a packet of qubits in the quantum channel. This handshake involves at least two packets: one transmission request (including the header) and one acknowledgement (ack), see Fig. \ref{fig:fig21transprop}. If there was an error in the received packet due to the channel or some multi-user collision, both entities must retransmit the header. Later, the sender can transmit the quantum packet to the receiver and the latter replies with an ack via the classical channel. 

The time delay for transmitting a request and ack packets for the $l$-th link in the classical channel can be computed as
\begin{equation}\label{eq:time_req}
	T_{\text{req},l} = T_{\text{ack},l} = \frac{T_{\text{bo},l}+T_{\text{tb}}+T_{\text{prop},l}}{P_{\text{sb}}} + T_{\text{proc},l},
\end{equation}
\begin{equation}\label{eq:time_tb}
	T_{\text{tb}} = N_{b} R_{b}, \quad T_{\text{prop},l} = c / d_{l}, \quad 1 \leq l \leq L = \eabsn{ \mathcal{L}_{c} \cap \mathcal{L}_{q}}{},
\end{equation}
where $L$ is the total number of links. $T_{\text{req},l}$ and $T_{\text{ack},l}$ denote the time delay for transmitting a request and ack packets at the $l$-th link, respectively. $T_{tb}$ is the transmission time of the packet, $N_{b}$ is the number of bits in a packet and $R_{b} \eds{\text{bps}}$ accounts for the transmission rate. It is assumed that both request and ack packets have the same length. 
$T_{\text{bo},l}$ is the back-off time for the $l$-th link which corresponds to the amount of time that the transmitter is waiting to perform a retransmission after detecting a collision. $ T_{\text{proc},l}$ and $ T_{\text{prop},l}$ correspond to the processing and propagation times of the $l$-th link, respectively, $c$ is the speed of the light and $d_{i}$ represents the distance of the $i$-th link.
$P_{\text{sb}}$ represents the probability of performing a successful transmission of an uncoded packet, given by
\begin{equation}\label{eq:prob_suc}
	P_{\text{sb}} = \ecs{1-P_{\text{col}}} \ecs{1-P_{b}}^{N_{b}},
\end{equation}
where $P_{b}$ corresponds to the error bit probability and $P_{\text{col}}$ is the probability of collision, for the case of a wired link $P_{\text{col}} = 0$. Uncoded packet is used for simplicity, in practice we need to take into account the error correction code.

After the handshake, the transmission of the qubits via the quantum channel is allowed, making it a contention-free channel. Then, the time delay for the transmission of a quantum packet ($T_{\text{tq}}$) can be obtained as
\begin{equation}\label{eq:time_txq}
	T_{\text{tq}} = \frac{N_{q} R_{q}+T_{\text{prop},l}}{1-P_{q,l}^{N_{q}}} + T_{\text{proc},l}, \quad 1 \leq l \leq L,
\end{equation}
where $N_{q}$ is the number of qubits in a packet, $R_{q} \eds{\text{qbps}}$ accounts for the quantum transmission rate, $P_{q,l}$ corresponds to the error qubit probability at the $l$-th link. Note that the denominator of (\ref{eq:time_txq}) is different as compared to (\ref{eq:time_req}) and (\ref{eq:prob_suc}) since, at least, one out of $N_{q}$ qubits in the packet is enough to perform the entanglement-swapping. Then, each receiver replies with an ack message via the classical channel. The average time delay of this proposal is lower than \cite{qwrap}, especially when the error probability is high ($P_{\text{sb}}$) which increases the retransmissions. In Fig. \ref{fig:fig21transref}, the header of bits and data of qubits are sequentially transmitted. If any packet is wrong, the sender must retransmit everything again increasing the overall time delay. Additionally, the receiver must synchronize both channels since the qubits may arrive later due to the different rates at both links, which is resource-demanding.

\subsection{Ethernet Header}

The Ethernet header lasts $20$ bytes, which is built by the MAC origin and destination addresses, Ether type, \acrfull{crc}, etc. To deploy the new QP over Ethernet, it is proposed to add a new option in the field of the Ether type as \textit{Ether Type = Quantum}. Once the Ethernet frame arrives at a new entity, this new option will notify that transmission over the quantum channel is witnessed. Meanwhile, the other fields of the Ethernet remain unmodified, since the MAC addresses are helping with the addressing and the CRC is verifying the integrity of the whole frame.

\subsection{Quantum Protocol Header}

The frame of the proposed QP is purely a header since the data are the qubits transmitted through the quantum channel. This protocol is connection-oriented since a virtual circuit must be first established before performing the end-to-end entanglement between two users. We propose that the header of the QP lasts for $20$ bytes and they are addressed as follows:
\begin{itemize}
	\item Sequence number ($4$ bytes): each frame transmitted must provide a sequence number to identify each packet and it also helps to sort the packets.
	\item Ack sequence number ($4$ bytes): This field is used to confirm the correct reception of a packet to the sender.
	\item Message type and Ack flag ($7$ + $1$ bits): the message type will be encoded in the first seven bits, and all these messages are detailed in the following sections. Meanwhile, the last bit represents the ack bit. 
	\item E2E entanglement identifier (ID) ($8$ bytes): a unique number that identifies each existing entanglement between two edge user devices. This will help the switches to build the entanglement table.
	\item Level number ($1$ byte): it is used to identify the progress of the entanglement-swapping. 
	\item Token ID ($2$ bytes): it is used only for transferring the token ID after performing a successful swapping.
\end{itemize}

\section{Link Establishment for Deploying Quantum Protocol over Ethernet}
\label{sec:estab}

The network executes essential protocols to continuously update switching tables, accurately mapping device addresses to network paths before transmitting both bits and qubits. These self-managing mechanisms are critical for ensuring a stable and responsive network environment, especially for the coordination required by quantum internet infrastructure.


\subsection{Quantum-Spanning Tree Protocol (Q-STP)}

Analogously to STP, the Q-STP is proposed to build an additional loop-free topology for the LAN of interest. However, the cost values must consider both classical and quantum channels, instead of exclusively accounting for the classical one in STP. The cost of the $l$-th link can be obtained as
\begin{equation}\label{eq:cost}
	C_{l} = C_{c,l} + C_{q,l}, \quad \forall l \in \mathcal{L}_{c} \cap \mathcal{L}_{q},
\end{equation}
where $C_{c,l}$ and $C_{q,l}$ are the cost of the $l$-th classical and quantum links, respectively, and $\mathcal{L}$ corresponds to the set that contains all the links in the LAN.

\eAddTwoColFig{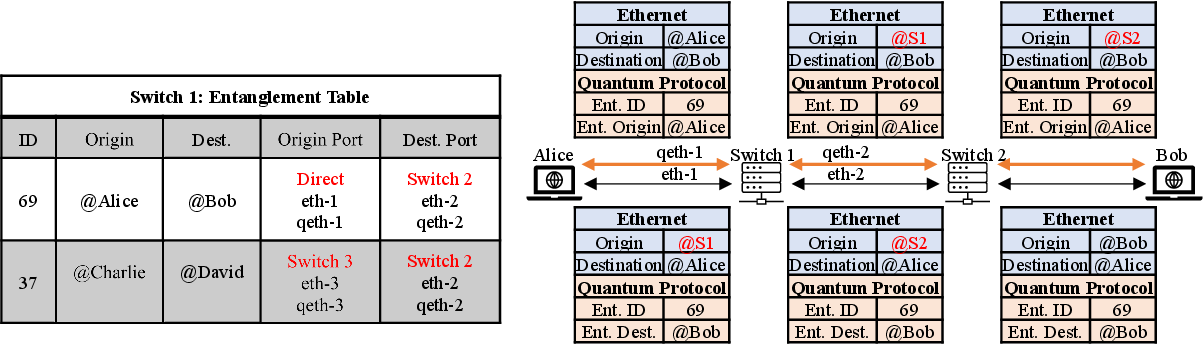}{0.93}{fig2circuit}{Example of the execution of the Path Establishment Protocol between Alice and Bob via two switches.}

\subsection{Discovery Protocol}



Before starting the transmission, the sender device should ensure that the receiver exists and is available to receive both bits and qubits. The explanation of this Discovery Protocol is given by exemplifying a particular case that Charlie would like to check the availably of David in Fig. \ref{fig1switches}. It is assumed that Charlie knows the MAC address of David, and Charlie sends a \textit{Discovery Request} message to the switch $\#1$ via the Ethernet port associated with a quantum port. Since the switch $\#1$ does not have any information about David, it will broadcast the \textit{Discovery Request} message through all those Ethernet ports attached to a quantum one. Finally, once David has received the \textit{Discovery Request} message, he will reply to Charlie with a \textit{Discovery Reply} message, and then, all the switches can update the information about Charlie and David. 


Making use of (\ref{eq:time_req}), the total delay for the discover protocol $T_{\text{disc}}$ can be given as
\begin{equation}\label{eq:time_disc}
	\begin{split}
		T_{\text{disc}} & = \sum_{l=1}^{L} \ecs{T_{\text{req},l}+T_{\text{ack},l}} \\
		& = 2L \frac{T_{\text{tb}}}{P_{\text{sb}}} + 2\sum_{l=1}^{L}  \frac{T_{\text{bo},l}+T_{\text{prop},l}}{P_{\text{sb}}} + T_{\text{proc},l}.
	\end{split}
\end{equation}
In the case that all the switches were equipped with the same hardware, the back-off time was the same value for all transmissions, and all links had the same length, (\ref{eq:time_disc}) could be simplified as
\begin{equation}\label{eq:time_disc2}
	T_{\text{disc}} = L\ecs{T_{\text{req}}+T_{\text{ack}}} = 2LT_{\text{req}},
\end{equation}
where the subscript $l$ was dropped out.

\subsection{Path Establishment Protocol}

The Ethernet network is self-configuring to real-time conditions by automatically adjusting paths based on current network demands. This adaptive behaviour optimizes performance, ensuring continuous connectivity with minimal manual intervention. The chosen path for each packet from a specific origin to its destination may differ due to this dynamic management. In parallel, maintaining end-to-end entanglement between two users necessitates a carefully coordinated sequence of entanglement-swapping processes across multiple intermediate switches. Given the adaptive nature of the network, it is essential to track which specific switches are engaged in this chain, as any alteration to the path must be promptly communicated to reconfigure a new chain of switches.

The Path Establishment Protocol can create a virtual circuit between the two users and allocate the required resources (classical and quantum interfaces) to perform the end-to-end entanglement. An illustrative example of this proposed protocol is depicted in Fig. \ref{fig2circuit}. Alice sends a \textit{Establishment Request} which is built by the entanglement ID and her MAC address, which corresponds to the entity that has initiated this process. Later, the switch $\#1$ is overwriting its MAC address in the Ethernet header and forwarding it to the next switch. Then, once the message has arrived to Bob, he transmits a \textit{Establishment Reply} back to Alice following the same procedure. By performing this MAC address manipulation, each intermediate switch knows which interfaces are forced to use for a given entanglement ID, and all this information is collected in an entanglement table. The established link between each pair of switches must be constantly monitored, and they must send a periodical control message to their neighbours. If nobody was replying, the switch should notify the edge users with a message of \textit{Establishment Interrupted} and restart the whole process again, if it is needed.

The total delay required by the Establishment protocol has, at least, the same time delay of the previous Discovery protocol ($T_{\text{est}} \geq T_{\text{disc}}$), given in (\ref{eq:time_disc}). Both protocols require the same number of exchanged messages in the network. However, the delay of the Establishment protocol may be larger than the Discovery one due to the processing time in the switch. The Establishment protocol not only requires building a new entanglement table, but it also has to manipulate the received messages before forwarding them. 

\eAddTwoColFig{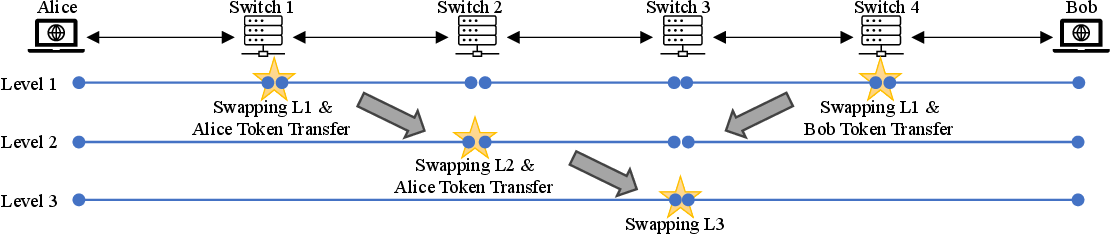}{0.93}{fig3swaptoken}{Illustrative example of the proposed sequential entanglement-swapping in a distributed network.}

\section{End-to-End Entanglement aided by Quantum Protocol over Ethernet}
\label{sec:qpeth}

After establishing a virtual circuit, the end-to-end entanglement procedure can commence. To simplify the explanation of the proposed protocol and procedures, we assume that Alice wishes to establish entanglement with Bob through multiple intermediary switches, and the Path Establishment Protocol is already completed. First, a sequential swapping strategy is explained to achieve progressive entanglement-swapping across the switches. The protocol is then presented in detail, accompanied by an analysis of the time delay involved. All aspects of the protocol are illustrated in Figures \ref{fig3swaptoken} and \ref{fig4swapmsg} to clarify each step of the process.

\subsection{Sequential Swapping in a Distributed Network}

The swapping process is typically managed by a central unit \cite{qrou1,qrou2,qrou3} in collaboration with the switches. However, the Ethernet network is a distributed one, and we propose a sequential swapping procedure starting from the two switches closest to each user, and the swapping is carried out step-by-step through the chain of switches established before, until the last switch which corresponds to that switch placed at the middle of the chain. To facilitate the explanation, an illustrative example of the proposed sequential swapping is given in Fig. \ref{fig3swaptoken}. Firstly, it is considered that all entities of the chain are entangled to their respective contiguous neighbours. Then, the switches $\#1$ and $\#4$ are two entities that are starting the swapping that involves the switches $\#2$ and $\#3$, respectively, and let us denote it as the \textit{level} 1 swapping and the probability of performing a successful swapping is denoted by $P_{\text{swap}}$. After succeeding in this \textit{level} 1 swapping, the switches $\#1$ and $\#4$ are not only notifying the success to the switches $\#2$ and $\#3$, respectively, but they are also sending their respective \textit{token} to them. By assuming that the switch $\#2$ is faster than switch $\#3$, the former is performing a \textit{level} 2 swapping and transferring its token to the latter. Finally, the switch $\#3$ is performing the \textit{level} 3 swapping. 

The concepts of \textit{level} and \textit{token} are introduced in this protocol to manage the swapping process, especially focusing on potential failures. There are two \textit{tokens} in the process which are generated at the two switches closest to both edge users. This \textit{token} is being transferred to the next switch in the chain once a successful swapping is carried out. In the case of swapping failure, the \textit{token} must either be sent back to or restarted at the two original switches closest to the edge users. Each \textit{token} must also account for the \textit{level}, which is the number of consecutive successful swapping processes for each direction of the chain. The highest \textit{level} number ($V$) for a given virtual circuit can be computed as
\begin{equation}\label{eq:levels}
	V = \efloor{S/2} + 1, \quad S \leq S_{t}
\end{equation}
where $S$ is the number of switches involved in the virtual circuit. In the case that a swapping fails at level $v \leq V$, its corresponding switch must notify about this failure to its one contiguous neighbour in the forward path and $v$ contiguous neighbours in the backward path. Hence, $v$ is also accounting the number of hops that the notification must reach. This swapping failure message is required since all the entities must regenerate the entangled bits and share them again. Finally, if a switch has both \textit{tokens}, then it means that it has to perform the last swapping or the virtual circuit, and it must also notify both users that the swapping process is completed.


\subsection{Swapping Protocol via Classical and Quantum Links}

To ease the description of the swapping protocol, an example is presented in Fig. \ref{fig4swapmsg}. After establishing the virtual path for the end-to-end entanglement, each switch must first generate some entangled qubits and share them with its neighbours. Before sharing these entangled qubits via the quantum channel, each pair of switches must perform a handshake protocol via the classical channel. Since Alice (left) was the user who started the process to entangle with Bob (right) in this particular case, it is assumed that all the entities involved are simultaneously transmitting the \textit{Point-to-point Entanglement Request} message from the left entity to the right entity, and then a \textit{Point-to-point Entanglement Reply} in the contrary way. The time delay of the entanglement request is given by
\begin{equation}\label{eq:time_er}
	T_{\text{er}} = \emaxs{l}{T_{\text{req},l}+T_{\text{ack},l}} = \emaxs{l}{2T_{\text{req},l}}, \quad 1 \leq l \leq L,
\end{equation}
Again, analogously to (\ref{eq:time_disc2}), by making the assumptions that all switches are made by the same hardware and all links have the same size, (\ref{eq:time_er}) can be simplified as $T_{\text{er}} = 2T_{\text{req}}$.

The time delay for transmitting the point-to-point entangled qubits ($T_{\text{ptp}}$) can be obtained as 
\begin{equation}\label{eq:time_ptp}
	T_{\text{ptp}} = \emaxs{l}{T_{\text{tq},l}+T_{\text{ack},l}}, \quad 1 \leq l \leq L.
\end{equation}
By adopting the same assumptions made for (\ref{eq:time_er}), (\ref{eq:time_ptp}) can be simplified as $T_{\text{ptp}} = T_{\text{tq}}+T_{\text{ack}}$.

After receiving the entangled qubits, those two switches possessing the \textit{token} start the swapping procedure. Again, a \textit{Swapping Request} and \textit{Swapping Reply} messages are needed to notify the following switch of the path that its entangled qubit is going to be swapped. If the swapping failed, a \textit{Swapping Error} message would be sent to its neighbours according to the \textit{level} number, as described in the previous subsection. After receiving the \textit{Error Ack} message, the point-to-point entanglement is restarted from the beginning. If the swapping succeeded, the \textit{token} and the success notification would be sent to the following switch by using the \textit{Token Transfer} and \textit{Token Ack} messages. Finally, once the last switch made the last swapping, it sends a \textit{Swapping Complete} and \textit{Complete Ack} messages to notify all the entities that the swapping procedure has successfully concluded.

\eAddTwoColFig{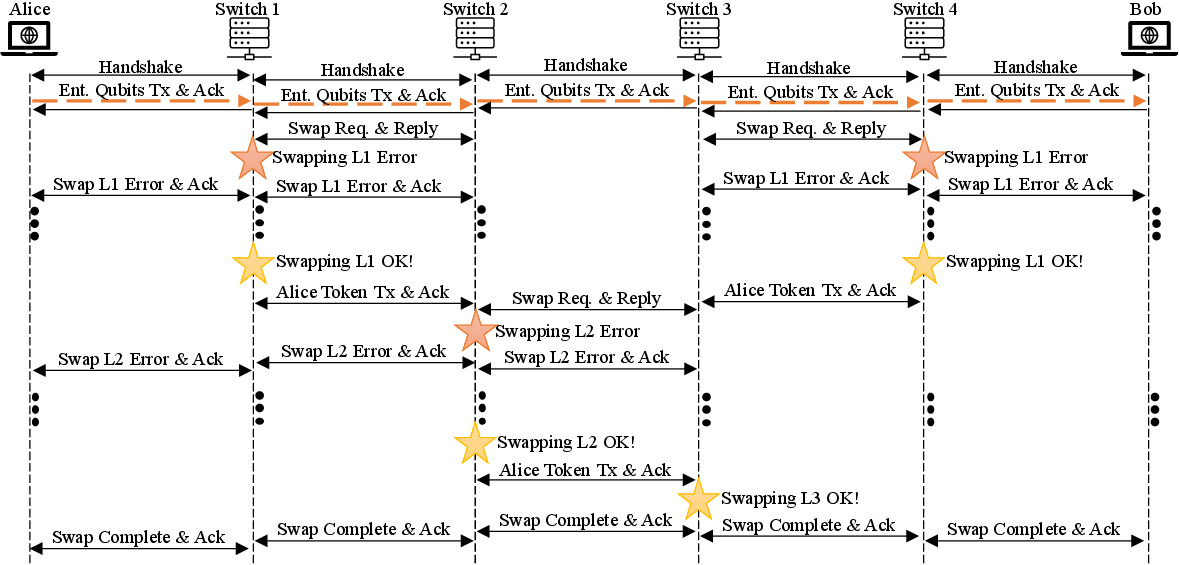}{0.93}{fig4swapmsg}{Illustrative example of the proposed swapping protocol by jointly using the classical and quantum links.}

The time delay for notifying a swapping error at the $v$-th \textit{level} ($T_{\text{error},v}$) is given by
\begin{equation}\label{eq:time_error}
	T_{\text{error},v} = \sum_{l=1}^{v}T_{\text{req},l}+T_{\text{ack},l}, \quad v \leq V.
\end{equation}
By making the same assumption of (\ref{eq:time_er}), (\ref{eq:time_error}) can be simplified as $T_{\text{error},v} = 2v T_{\text{req}}$. The time delay for the notifying that the swapping has completed ($T_{\text{comp}}$) is a particular case of (\ref{eq:time_error}) by substituting $v=V$. The time delay of the other remaining messages corresponds to a standard request and reply given by (\ref{eq:time_req})-(\ref{eq:prob_suc}). 

\subsection{Analysis of the Swapping Protocol Time Delay }
All the time delays given in the previous subsections only describe the amount of time required for sending each particular message. However, since the entanglement-swapping procedure may fail, all the described protocols must be repeatedly executed until all the swapping processes are successfully performed at each switch of the virtual circuit. Hence, the total required time delay for the $(T_{\text{swap}})$ is obtained by considering the swapping error probability ($P_{\text{swap}}$). To ease the notation, it is assumed that all switches are built with the same hardware and all links are the same size.

On the one hand, in the particular cases where there is either a direct connection ($S=0$) or one single switch ($S=1$) between the two edge users, the total time delay for swapping is given as
\begin{equation}\label{eq:time_swap0}
	T_{\text{swap}} = T_{\text{er}} + T_{\text{ptp}} = 3T_{\text{req}} + T_{\text{tq}},
\end{equation}
which corresponds to the transmission of the point-to-point entanglement request and the entanglement qubits. For the former case ($S=0$), the user who has taken the initiative of establishing an entanglement is responsible for sharing the entangled qubits. Meanwhile, for the latter case ($S=1$), once the sender has requested the point-to-point entanglement process, the switch is responsible for sharing the entanglement qubits between the two edge users, and therefore, reducing the required overhead.

On the other hand, when two or more switches are interconnecting the two edge users ($S\geq 2$), the total time delay for swapping ($T_{\text{swap}}$) can be computed as
\begin{equation}\label{eq:time_swap1}
	\begin{split}
		T_{\text{swap}} & =  \ecs{T_{\text{tq}}+T_{\text{ack}}}N_{\text{t}}^{V} + \ecs{T_{\text{req}}+T_{\text{ack}}} \\
		& \times \ecs{3N_{\text{t}}^{V}+2N_{\text{t}}^{V-1}} \sum_{v=1}^{V-2}(v+1)N_{\text{t}}^{V-v-1},
	\end{split}
\end{equation}
where $N_{t} = P_{\text{swap}}^{-1}$ is the average number of trials to achieve a successful entanglement-swapping. By making use of (\ref{eq:time_req}) and (\ref{eq:time_txq}), (\ref{eq:time_swap1}) can be simplified as
\begin{equation}\label{eq:time_swap2}
	\begin{split}
		T_{\text{swap}} & = T_{\text{tq}}N_{\text{t}}^{V} + T_{\text{req}} \ecs{7N_{\text{t}}^{V} + 4N_{\text{t}}^{V-1}} \\
		& + 2T_{\text{req}}\sum_{v=1}^{V-2}(v+1)N_{\text{t}}^{V-v-1},
	\end{split}
\end{equation}
where the first term of (\ref{eq:time_swap2}) corresponds to the time delay devoted to the quantum channel, meanwhile the rest of the terms correspond to the time delay required in the transmission at the classical one. Note that the former can be the dominant term if the length of the link is significantly large since the qubit error probability is strongly enhanced, and hence, the QSTP must consider the cost of both links.

\section{Numerical Results}
\label{sec:num}

In this section, a numerical assessment of the proposal QP is given in terms of time delay in a realistic network size. The proposed handshake protocol to coordinate the transmission over the quantum channel is compared to the existing solution \cite{qwrap}, which is the baseline case.

It is considered that the quantum channel is deployed in a wireless link and qubit error probability is given in \cite{airchan}.
In Fig. \ref{fig:handshake}, a comparison between the proposed handshake and the existing \cite{qwrap} is provided for different values of link length $l$ and $P_{\text{col}}=0.1$ and $0.3$. The qubits are assumed to be immediately transmitted via the quantum channel after the Ethernet header is sent through the classical channel for the baseline. The performance of the proposal is always better than the baseline for all values of link length and probability of collision. Our proposal is better since the classical and quantum packets are independently managed, in the case that there was either a collision or an error in the classical packet, the transmitted would only transmit the QP packet again, instead of transmitting both classical and quantum packets in \cite{qwrap}. Then, once both entities have agreed, the quantum packet can be transmitted which does not suffer from any collision.

In Fig. \ref{fig:distdelay}, the performance in terms of time delay for the proposed QP is provided. Firstly, let us define $D$ as the total distance between the two edge users. Then, the number of required switches to interconnect them can be computed as $S = \eceil{D/l}$. Again, it can be seen that when the link length is increased ($l\uparrow$), the time delay is increased since the path loss of the optic fibre is also enhanced. However, increasing the length of the link may imply a reduction in the number of required switches ($S\downarrow$), and it strongly reduces the time delay since the number of entanglement-swapping is also lowered. Moreover, to keep reducing the time delay of long link lengths, more than one entangled qubit can be sent in each quantum packet to reduce the required retransmission in the quantum channel. In this case, the qubits are increased to $N_{q} = 100$ qubits, and this effect is especially relevant for long link lengths and long distances ($l\geq 70$ km, $D=400$ km). According to \cite{400km}, photon transmissions through optical fibre have proven feasible over long distances ($D\sim 421$ km), meanwhile all of our simulations operate exclusively within this distance for link lengths ($l\leq 400$ km).


\eAddFig{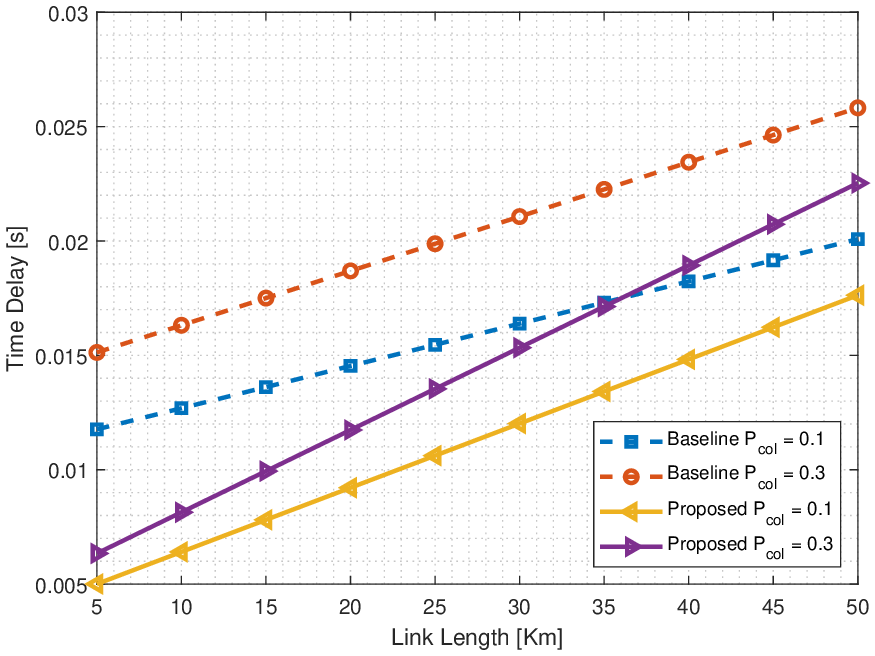}{0.95}{fig:handshake}{Comparison in terms of time delay between the proposed technique and \cite{qwrap} for $P_{\text{col}}=0.1$ and $0.3$ and different values of link length $l$.}

\eAddFig{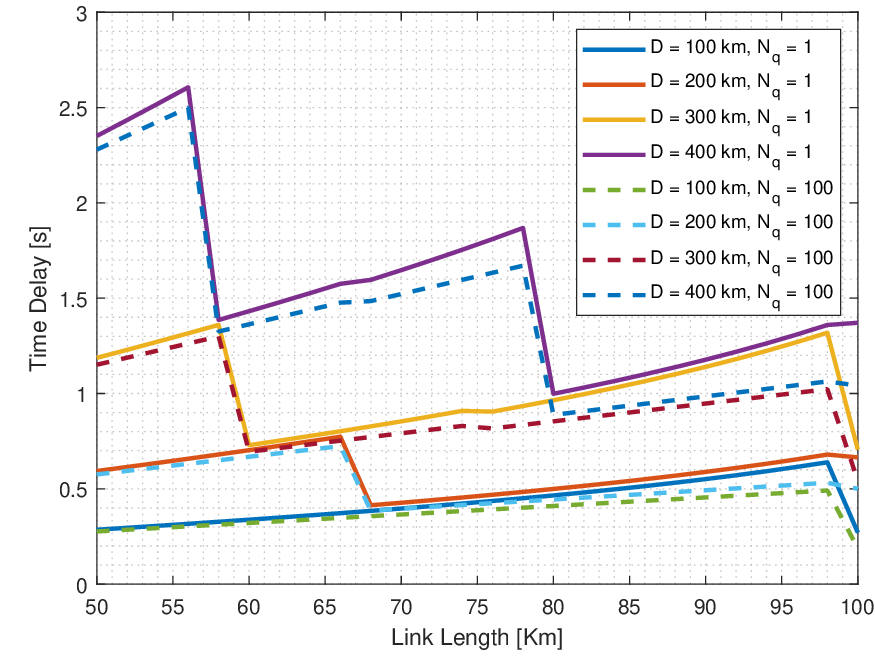}{0.95}{fig:distdelay}{Performance of the proposed QP in terms of time delay for different values of the total distance between the users ($D$) and link lengths ($l$).}

\eAddFig{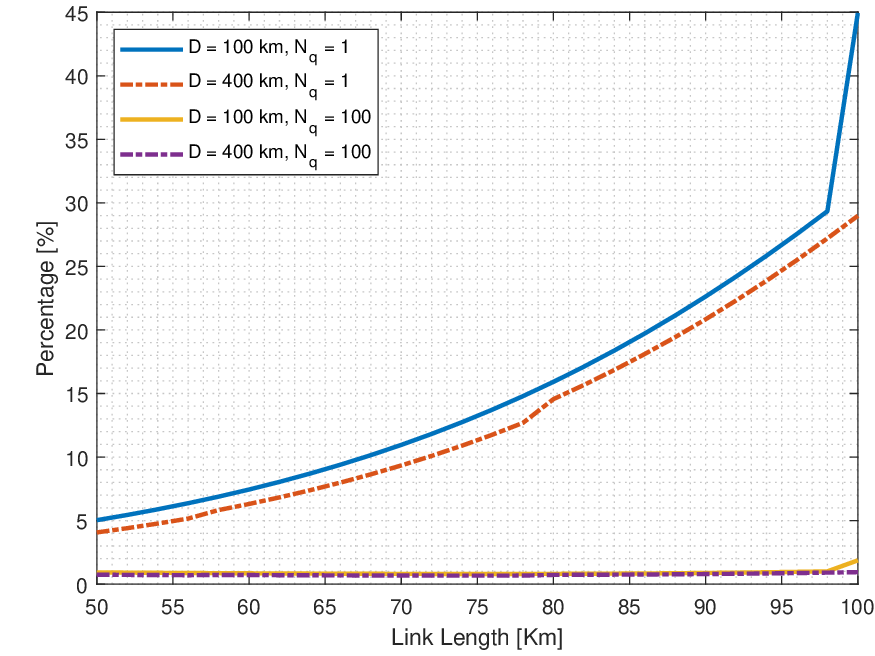}{0.95}{fig:percentage}{Performance of the proposed QP in terms of percentage of time delay devoted to quantum channel over the total one for different values of the total distance between the users ($D$) and link lengths ($l$).}

In Fig. \ref{fig:percentage}, the ratio between the required time delay for transmitting in the quantum channel over the total value given in Fig.~\ref{fig:percentage} is presented. For the case that we are only sending one entangled qubit ($N_{q}=1$), the time delay required is significantly raised as the link length is increased since more retransmissions are needed due to the strong attenuation and scattering. However, when more entangled qubits are sent in the quantum packet ($N_{q}=100$), the time delay of the quantum channel is strongly reduced since the probability of receiving, at least, only one correct for the entanglement-swapping is significantly increased.

\section{Conclusions}
\label{sec:conclusion}

This research demonstrates the potential of deploying QP over Ethernet networks, highlighting the synergy between classical and quantum technologies. By leveraging key Ethernet features, such as addressing, MAC, and scalability, this approach presents a practical way to support the complex requirements of quantum communication, including precise coordination and efficient switching. The novel protocols introduced in this work effectively achieve end-to-end entanglement, and the analysis of time delay for each protocol shows that our solution outperforms existing approaches in the literature, with total delay remaining well below the decoherence time, ensuring stable and practical implementation in real-world environments. Ultimately, our approach not only addresses the immediate demands of quantum networking but also establishes a foundational framework for future advancements in secure data exchange.


\ifCLASSOPTIONcaptionsoff
\newpage
\fi
\bibliographystyle{IEEEtran}
\bibliography{./bibtex/IEEEabrv,./bibtex/IEEEexample}{}

\end{document}